# PROSPECTS FOR P$^{11}$B FUSION WITH THE DENSE PLASMA FOCUS: NEW RESULTS

Eric J. Lerner

Lawrenceville Plasma Physics
9 Tower Place
Lawrenceville, NJ 08648, USA

## ABSTRACT

Fusion with p$^{11}$B has many advantages, including the almost complete lack of radioactivity and the possibility of direct conversion of charged particle energy to electricity, without expensive steam turbines and generators. But two major challenges must be overcome to achieve this goal: obtaining average ion energies well above 100keV and minimizing losses by bremsstrahlung x-rays. Recent experimental and theoretical work indicates that these challenges may be overcome with the dense plasma focus.

DPF experiments at Texas A&M University have demonstrated ion and electron average energies above 100keV in several-micron-sized hot-spots or plasmoids. These had density-confinement-time-energy products as high as 5.0 x10$^{15}$ keVsec/cm$^3$. In these experiments we clearly distinguished between x-rays coming from the hot-spots and the harder radiation coming from electron beam collisions with the anode.

In addition, new theoretical work shows that extremely high magnetic fields, which appear achievable in DPF plasmoids, will strongly reduce collisional energy transfer from ions to electrons. This reduction has been studied in the context of neutron stars and occurs when ion velocities are too small to efficiently excite electron transitions between Landau levels. It becomes a major effect for fields above 5 gigagauss. This effect will allow average electron energies to stay far below average ion energies and will thus reduce x-ray cooling of p$^{11}$B. In this case, fusion power will very significantly exceed x-ray emitted power. While fields of only 0.4 gigagauss have so far been demonstrated with the DPF, scaling laws indicate that much higher fields can be reached.

These theoretical predictions need to be tested in new sets of experiments. The efficiency of energy transfer into the plasmoids must also be optimized. Fortunately the compact and inexpensive nature of the DPF means that such work can be carried out with modest resources. If successful, this line of research can lead to power production with costs more than an order of magnitude below those of current energy sources.

## 1. INTRODUCTION

Controlled fusion with advanced fuels, especially hydrogen-boron-11, is an extremely attractive potential energy source. Hydrogen-boron fuel generates nearly all its energy in the form of charged particles, not neutrons, thus minimizing or eliminating induced radioactivity. The main reaction, p+$^{11}$B-> 3$^4$He, produces only charged particles. A secondary reaction, $^4$He+$^{11}$B-> $^{14}$N +n does produce some neutrons as the alpha particles produced by the main reaction slow down in the plasma, but only about 0.2% of the total fusion energy is carried by the neutrons, whose typical energy is only 2.5 MeV. Hydrogen-boron fuel also allows direct conversion of charged-particle energy to electric power, without the expensive intermediate step of generating steam for turbines.[1-3]

However, fusion with such fuels requires average ion energies above 100 keV (equivalent to 1.1 billion K) in a dense plasma. Experiments reported here have met this requirement and demonstrated electron and ion energies over 100 keV in a dense plasma focus device which is compact and inexpensive. This was achieved in plasma "hot spots" or plasmoids that, in the best results, had a density-confinement-time-energy product of $5.0 \times 10^{15}$ keVsec/cm$^3$, a record for any fusion experiment and a factor of ten above the best achieved in the much larger Tokamak experiments.[4] To do this, the experiment used a higher-than-usual operating pressure for the device, which increased particle energy. An X-ray detector instrument measured the electron energies in a way that eliminated previously-existing doubts that the hard X-rays were generated by the hot spots.

The dense plasma focus (DPF), first invented in 1954, is far more compact and economical than other controlled-fusion devices. It consists of two coaxial cylindrical electrodes usually less than 30 cm in all dimensions in a gas-filled vacuum chamber connected to a capacitor bank. The total cost of such a device is under $500,000, for a 500 kJ system. It is capable of producing high-energy X-ray and gamma-ray radiation and intense beams of electrons and ions, as well as abundant fusion reactions.[5] In operation, the capacitors discharge in a several-microsecond pulse, the gas is ionized and a current sheath, consisting of pinched current filaments, forms and runs down the electrodes. When the sheath reaches the end of the inner electrode (the anode), the filaments pinch together, forming dense, magnetically-confined, hot spots or plasmoids.[6-8] The plasmoids emit soft X-rays with energy in the range of several kiloelectron volts. X-ray pinhole images have demonstrated that the plasmoids are tiny, with radii of a few microns to tens of microns.[9-10] The plasmoids have densities in the range of $10^{20} - 10^{21}$ /cm$^3$. These densities have been measured in a number of independent methods including heavy ion fusion[11], $CO_2$ laser scattering[12], and x-ray line intensities[13]. These plasmoids emit intense beams of accelerated ions and electrons[14-15]. Fusion neutrons are emitted from the device in large quantities (up to $10^{13}$)

The role of the plasmoids in producing the fusion neutrons and the physical processes involved in their formation and maintenance have been hotly debated among DPF researchers for decades. The model that best fits all the existing data makes the role of the plasmoids central to neutron production. This model, initially developed by Bostick and Nardi[5], and confirmed by observations of several groups over three decades, was elaborated into a more quantitative theory by the present author[16,17]. In this model, the electron beam transfers part of its energy to the plasmoid electrons, which generate X-rays through collisions with nuclei. Through a plasma instability (probably ion-acoustic), the electrons then transfer part of their energy to the ions, with a typical delay (in our experiments) of ~10 ns. Ion collisions, generating fusion reactions and neutrons, then occur [18]. When the ion and electron beams have exhausted the magnetic energy that confines the plasmoid, and partially or wholly evacuated the particles in the plasmoid, the fusion reactions end.

The DPF routinely produces hard X-rays and gamma rays indicating the presence of bremsstrahlung radiation from high-energy electrons colliding with nuclei[17]. Together with independent evidence, this indicated that the hot spots contained ions and electrons at very high energies in the range of interest for advanced fuel fusion [11,16,17,18,20]. However, most researchers in the field have believed that the hard radiation was generated not in the plasma itself but by the electron beam when it strikes the end of the anode, and that the hot spot plasma was relatively cool with electron energies of only several kiloelectron volts, too low for hydrogen-boron fusion[21]. This key question of where the hard x-rays came from remained unresolved by previous experiments.

## 2. EXPERIMENTAL SET-UP

To clearly distinguish between the radiation from the anode end and that from the plasmoids, we used a cylindrical anode with 6-cm-deep recess at the pinch end (Fig.1). In this manner, the point where the electrons hit the anode, generating gamma rays, would be separated in space from the plasmoids, which are located at the plasma pinch very close to the end of the anode. We blocked the line of sight between the bottom of the hole and a set of X-ray detectors with a 5-cm-thick lead brick. This reduced the intensity of 1 Mev gamma rays by 50-fold. The line of sight from the plasmoids to the x-ray detectors, in contrast, passed through a 1-mm Be window, which allowed all but the lowest energy X-rays (below 1 keV) to pass.

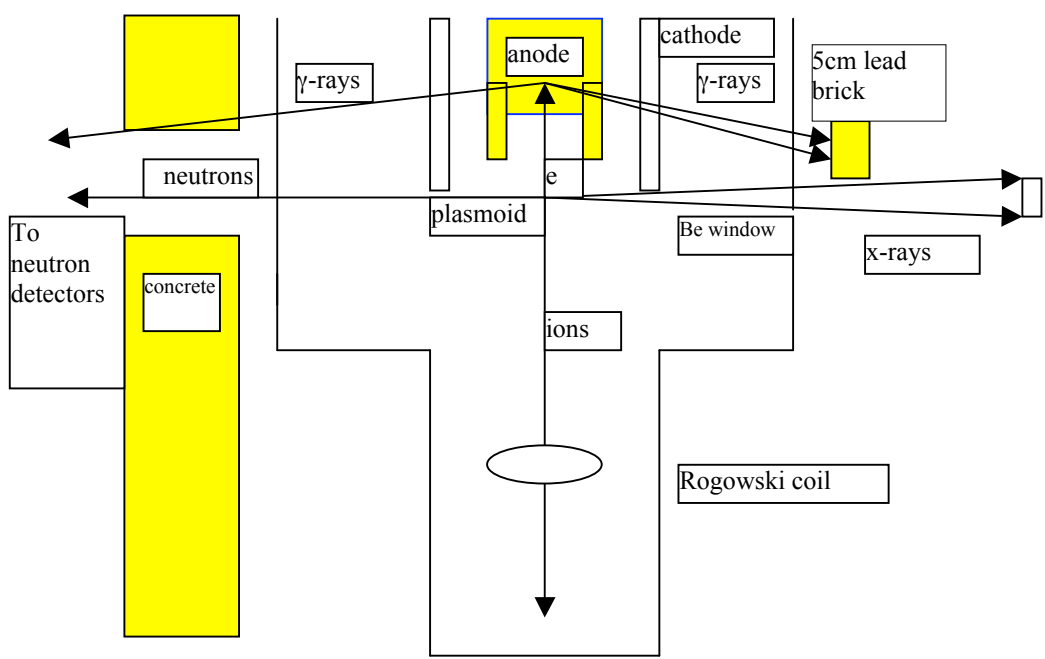

Fig. 1. In a dense plasma focus, a small plasmoid or hotspot forms near the end of the anode, emitting an electron beam towards the anode, an ion beam away from the anode as well as x-rays and neutrons from fusion reactions. X-rays from the plasmoid pass through a 1 mm thick Be window to the x-ray detectors, but gamma rays from collisions of the electron beam with the anode are blocked by a 5 cm thick lead brick. Neutrons pass through apertures in a 60-cm concrete shield wall to neutron detectors at 9 m and 17.4 m. Lines of sight of the two neutron detectors are ~90 degrees apart in azimuthal direction. A Rogowski coil at 65 cm from the anode measures the ion beam.

The experiments were performed at the Texas A&M University DPF facility that used a 268 microfarad capacitor bank, charged to either 30 or 35 kV. A shorter anode than is customary was used to increase the fill pressure which theoretical models indicated would increase the plasmoid density and particle energy[16,17]. For a given capacitor bank rise time, and a given anode radius, the length of the anode determines the optimum fill pressure. This is because the current must peak at the time the plasma sheath arrives at the end of the anode for a good pinch to occur. The velocity of the sheath in turn decreases with increasing fill pressure. So the longer the anode, the lower the optimal pressure.

Eric J. Lerner

However, if the anode radius is reduced, increasing the B field and sheath velocity, the same fill pressure corresponds to a longer anode.

With the existing experimental apparatus, it was not practical to change the anode radius, which was fixed at 5.1 cm prior to this experiment. To increase the fill pressure, the anode length was made as short as possible while still getting a good pinch. The fill pressure for each gas mix was optimized by maximizing either neutron yield or, for He, maximum dI/dt in the pinch.

The cathode radius was 8.6 cm, anode radius 5.1 cm, anode length 24 cm, and insulator length 7.5 cm. A slotted knife-edge around the insulator base was employed to enable operation at the higher pressure. Shots were fired using fill gases of deuterium, helium or helium-deuterium mixtures at a 50-50 ratio by ion number. At 30 kV, optimal fill pressure was 13 torr for D, 14 torr for He-D and 16 torr for He. Peak current was typically 1.2-1.3 MA at 30 kV and 1.4-1.5 MA at 35 kV. Helium was used as a fill gas because its μ and Z of 2 and 4 are relatively close to the average μ and Z of decaborane, the likely fuel for hydrogen-boron reactions (μ =2.66, z=5.17).

To measure the energy of the X-rays and infer the electron energy, 300-micron, 3-mm and 6-mm thick copper filters were placed in front of the three X-ray detectors (NE102 scintillators with photomultiplier tubes (PMT)). The ratio of X-ray intensity at a given instant recorded behind the 6mm filter to that recorded behind the 300 micron filter provided a measure of average X-ray energy, since only the more energetic X-rays could penetrate the thicker filter. Average electron energy $T_e$ was calculated by comparing the ratio of the 6 mm/300 micron-filtered signals observed to that calculated for bremsstrahlung emitted by Maxwellian distributions of electrons at different electron temperatures. The ratio of the 3-mm signal to the 300-micron signal was used to test if the energy distribution was Maxwellian or non-Maxwellian. Alternatively, calculated electron energy was also on the basis of assuming monoenergetic electrons. Remote scintillators and PMTs at 9.0 and 17.4 meters that were shielded by 5-cm thick lead bricks measured the gamma ray pulses from the anode end, and the neutrons produced by deuterium fusion.

To eliminate the possibility that radiation from off-axis sources passed through the chamber wall, in later shots the sides of the detectors were also blocked with lead bricks. However, results from earlier shots were very similar to those from later shots, so all data was used. RF noise from the pinch was suppressed with aluminum foil shielding, but still interfered with the X-ray signals on some shots. The following analysis utilized only the shots with a S/N ratio of more than 8 to 1. Shots with lower S/N had higher $T_e$, so their exclusion does not affect the conclusions presented here. In the D and D-He shots, confusion of X-ray and neutron signals was avoided since the neutrons reached the detectors 140ns after the X-ray signals and were relatively small due to the 1-2 mm thickness of the X-ray scintillators.

The remote scintillators were calibrated for neutron detection against calibrated silver activation neutron counters. Their sensitivity to gamma rays and x-rays was calculated based on data in the literature that related neutron to X-ray and gamma-ray sensitivity[22-23]. They were cross calibrated with the X-ray detectors (by measuring output from the same shot with all lead bricks removed), providing an *absolute* calibration of the X-ray detectors. The X-ray detectors were cross-calibrated with each other in several shots with the filters removed. Relative calibration measurements were consistent to within 10%. The accuracy of the absolute neutron calibrations, and thus of the absolute X-ray calibrations, is estimated to be better than 30%.

## 3. RESULTS

### 3.1 ELECTRON ENERGY

In all but one of the eight shots with highest signal-to-noise (S/N) ratio, average electron energy $T_e$ exceeded 100 keV and peak $T_e$ generally exceeded 150 keV. These are $T_e$ calculated on the Maxwellian assumption. (Table 1 shows the range of values) (Calibration uncertainties of ±10% imply ±20% errors in $T_e$.) A high $T_e$ is required for the high ion energy needed for hydrogen-boron fusion, since at the very high densities required for net energy production collisions with electrons with average $T_e$ below 100 keV will cool ions too rapidly (although high B fields will alter this situation as shown in section 5.). By comparison, $T_e$ in Tokamak Fusion Test Reactor experiments was, at most, 11.5 keV.

Table 1

| Fill gas | D | D-He | He |
|---|---|---|---|
| x-ray pulse duration, ns | 35 | 30 | 25 |
| Average, $T_e$ keV | 140-215 | 120-190 | 80-150 |
| Peak, $T_e$ keV | 210-500 | 180-250 | 100-210 |
| Neutronsx10$^9$ | 10-30 | 1-4 | |

The good agreement between peak $T_e$ calculated from the ratio of 6mm/300 microns and that calculated from the ratio of 3 mm/300 microns shows that the results are consistent with a Maxwellian distribution. (Fig.2) However, they are also consistent within calibration errors with monoenergetic electrons. In this case, average $T_e$ would range from 180 keV to 300 keV rather than from 80 keV to 215 keV for Maxwellian distributions.

The lead shields distinguished the X-rays generated by the plasmoids from the gamma rays coming from the anode tip. This demonstrated that we were indeed measuring electron energy in electrons trapped in the plasmoids, not just in the electrons in the beam. As measured by the remote scintillators, the gamma ray pulse signals were between 60 and 180 times smaller than the x-ray peaks so they could not affect the x-ray results. (Figs.3,4) For many of the shots, including all the He shots, the gamma ray pulses were not observable at all with the shielded detectors. In addition, the gamma ray pulses did not resemble the x-ray pulses, being 4-6 times shorter in duration (FWHM) with rise times generally 15 times faster. Finally, the fact that the X-ray pulses were not detectable when filtered by the lead brick confirms that the X-ray energy is concentrated below 600 keV, in contrast to the more than 1 MeV gamma rays from the anode.

There was a tendency for $T_e$ to decrease gradually with increasing Z in the fill gas. This is to be expected if the efficiency of energy transfer from the electron beam to the plasmoid electrons is constant across the gases as is the ratio of total beam energy to total particle number in the plasmoids. In this case, the increase in the ratio of electron number to total particle number means that the same energy is spread over a greater number of electrons, leading to a decrease in $T_e \sim (z+1)/z$. This is consistent with the observations summarized in table 1 and would extrapolate to a range of between 75-140 keV for comparable conditions and good pinches. This is well in the range of interest for p$^{11}$B reactions.

**3.2 PLASMA DENSITY**

Eric J. Lerner

These results show clearly that the high energy X-rays are emitted from high-energy electrons within the plasma. But do they come from the high-density plasmoids or do they come from a larger volume at lower density?

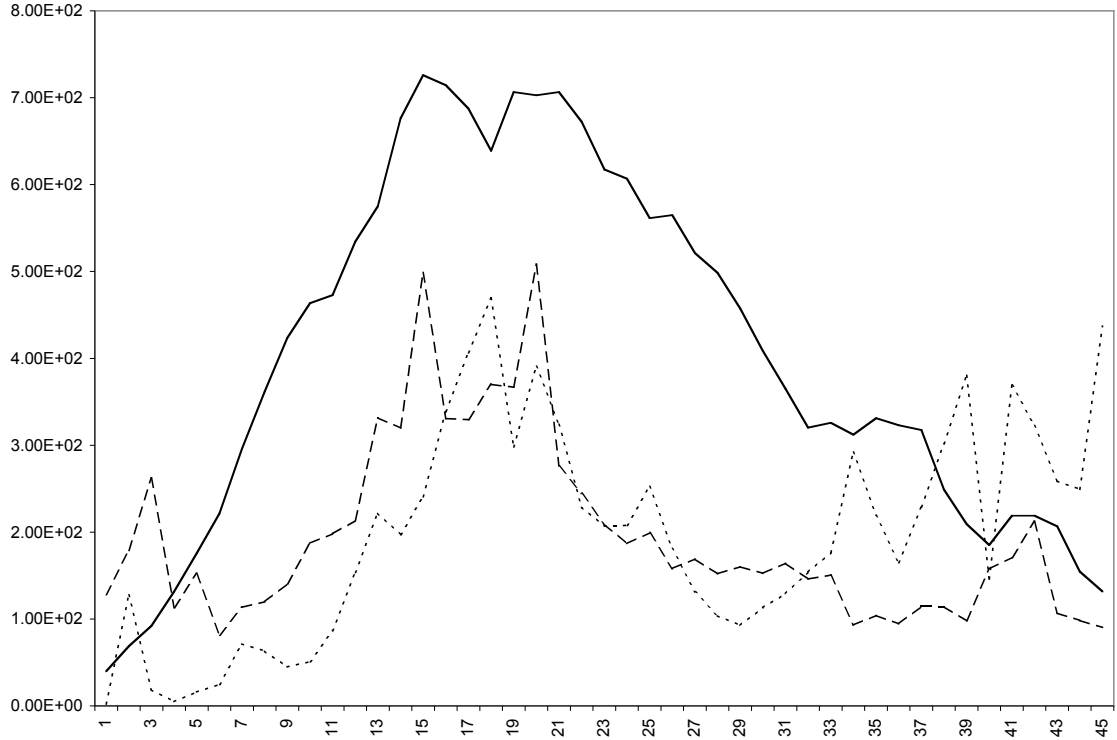

Fig. 2. X-ray power output(solid line), electron average energy $T_e$ calculated from ratio of 6mm/300 micron-filtered output(dashed), $T_e$ from ratio of 3mm/330 micron filtered output (dotted) for a single D shot(shot 81705, 35 kV, 15 torr, $9 \times 10^9$ neutrons). $T_e$ is in keV, while x-ray output, measured by 300 micron filtered detector, is in units of 350 W total emitted power. Time unit is 2ns. The close correspondence of the two measures of $T_e$ in the latter half of the pulse is consistent with a Maxwellian electron distribution, while the higher 6mm/300 micron $T_e$ in the earlier part of the pulse indicates a broader-than Maxwellian energy distribution. Average $T_e$ for this pulse is 200keV, and $T_I$, derived from neutron time of flight measurement, is 300keV. Note that x-ray pulse duration and the overall X-ray energy emitted is considerably greater in this shot at 35kV, with peak current of 1.5 MA than in Fig. 3, at 30kV and 1.3MA, showing rapid scaling of X-ray output.

We have determined the density of the plasma that produces the neutrons by the DD reaction. There is strong evidence that this is the same volume that traps the high-energy electrons and produces the hard X-ray radiation (see Sect. 3.4). In 25% of the deuterium shots at optimum fill pressure (two out of eight), 14.1 MeV neutrons were observed. These 14.1 MeV neutrons are produced by the fusion of deuterium nuclei with tritons produced by DD fusion reactions and then trapped in the high-density regions by strong magnetic fields. This was observed in one of the two deuterium shots with the best X-ray data (shot 81602), as well as in one shot (shot 80910) that was made before the X-ray detectors were functional(Fig.5). Shot 80910 also had the highest number of DD neutrons($5\times10^{10}$).

The DT neutrons were detected by time of flight measurements using the 9 meter and 17.4 meter scintillators. The DT neutron detection cannot possibly be cosmic rays or other noise sources. The delay time between the DT neutron peak at the 9 meter detector and that at the 17.4 meter detector is identical (with the 2 ns measurement window) to that for 14.1 MeV neutrons in shot 81602 and differs by only 4 ns (1%) for shot 80910 (Fig.5). In both shots, the FWHM length of the DT pulse, $\tau$, is 10 ns, significantly longer than the minimum width set by instrumental response times. (The recovery time of the scintillator and PMT is <1 ns, but since data is taken every 2 ns, the minimum pulse width observable is 4ns and still well under the 10 ns recorded.) There are no other such pulses at unrelated times (except for the much earlier gamma ray pulse). While the number of neutrons actually detected is not large (7 for shot 81602 and 3 for shot 80910), even a single 14.1 MeV neutron produced a pulse four times the instrument noise level and for shot 81602 the peak is more than 25 times larger than the noise level (Fig.6). The isotropic distribution of the neutrons is indicated by their detection at both detectors, which are 90 degrees apart in azimuthal direction.

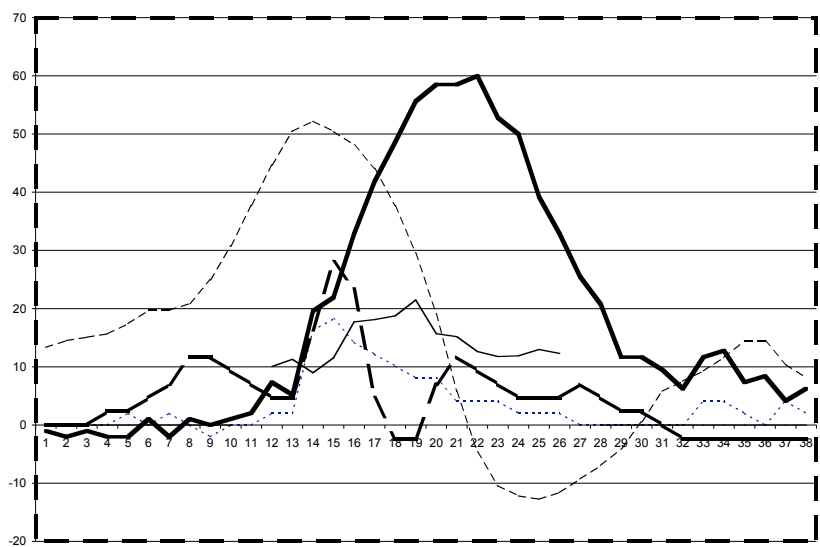

Fig. 3. dI/dt of device current(light dashed), x-ray output(heavy solid), ion beam current (heavy dashed), $T_e$ from 6 mm/300 micron ratio (light solid), gamma-ray output through 5cm lead shield of remote PMT(dotted) from a single D shot (Shot 81602, 30 kV, 12 torr, $1.0 \times 10^{10}$ neutrons). Units: dI/dt in 100 A/ns (inverted); x-ray output in 2 kW; ion beam current in 5 A; $T_e$ in 10 keV; gamma ray output in 20 W of power through the shield. All traces are referred to time of radiation from the pinch region. The shielded gamma ray pulse is far smaller than the unshielded x-ray pulses and differs from it in shape and timing. The gamma ray pulse generated by the electron beam hitting the anode and the ion beam are clearly correlated as both are with the dI/dt pinch. Heating of the electrons in the plasmoid continues as the beam tails off, reaching a peak 14 ns later.

In these two shots, the ratio of the number of 14.1 MeV neutrons to the number of 2.45 MeV neutrons provides a direct measure of density. In DD plasma approximately 0.9 tritons are produced by the D+D$\Rightarrow$T+p reaction for every 2.45 MeV neutron produced by the D+D⇒3He+n reaction. Thus the ratio of 14.1 Mev neutrons to 2.45 MeV neutrons is R = 0.9F, where F is the fraction of tritons that undergo fusion. Since

$$F = \langle\sigma v\rangle n_i \tau,$$

where $\langle\sigma v\rangle$ is the average product of the fusion cross section and the triton Maxwellian velocity distribution, $n_i$ is the deuteron plasma density and $\tau$ is the confinement time for the

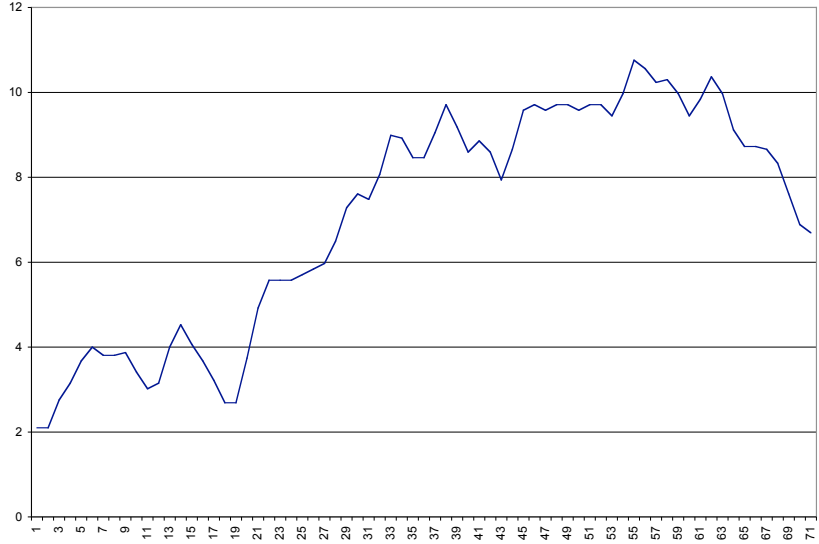

tritons, therefore

$$n_i = F/\langle\sigma v\rangle\tau.$$

Fig. 4 Energy transferred to the ions leads to neutron emission (kW of total neutron power). The broadening of the neutron peak, due to the energy spread of the neutrons, has been compensated for, although the peak is probably somewhat shorter in duration than portrayed here. (Shot 81602, 30 kV, 12 torr, 1.0 x$10^{10}$ neutrons).

The very short DT pulse duration, 10 ns, implies a small spread in triton velocity relative to the axis of the DPF. (The lines of sight to the neutron detectors are perpendicular to that axis.) This is consistent with the model of the plasmoid developed by Bostick, Nardi and other researchers and confirmed by x-ray pinhole images[10] in which the plasmoid has an "apple core" structure aligned with the DPF axis. Tritons contained by the strong magnetic fields within the plasmoid encounter the highest density plasma, and thus the most chance for fusion reactions while traveling along the plasmoid axis, and much lower density while re-circulating through the outer region of the plasmoid.

In shot 81602, F = 6.6±1x$10^{-3}$ and for shot 80910, F = 5.4±1.8x$10^{-4}$. The shot-to-shot variability is typical of DPF functioning without optimization, which was not possible for this set of experiments. We calculated n on the assumption that the tritons did not slow down significantly from their initial 1.01 MeV energy during the confinement time, and that DT neutron production was terminated as the plasmoid density decayed. Even at a density of 3.3x$10^{21}$/cm$^3$, the energy loss time for tritons in ion-ion collisions is 55 ns, much longer than $\tau$, supporting the assumption that the tritons have not slowed substantially by the end of the pulse. The energy loss time for the tritons at the same density in collisions with electrons at 100 keV energy is longer still, 350 ns. (A beam-plasma interaction that slows the tritons collectively cannot be entirely ruled out, although we believe it unlikely. If the tritons are slowed down to low energies during the 10 ns pulse, the reaction rate

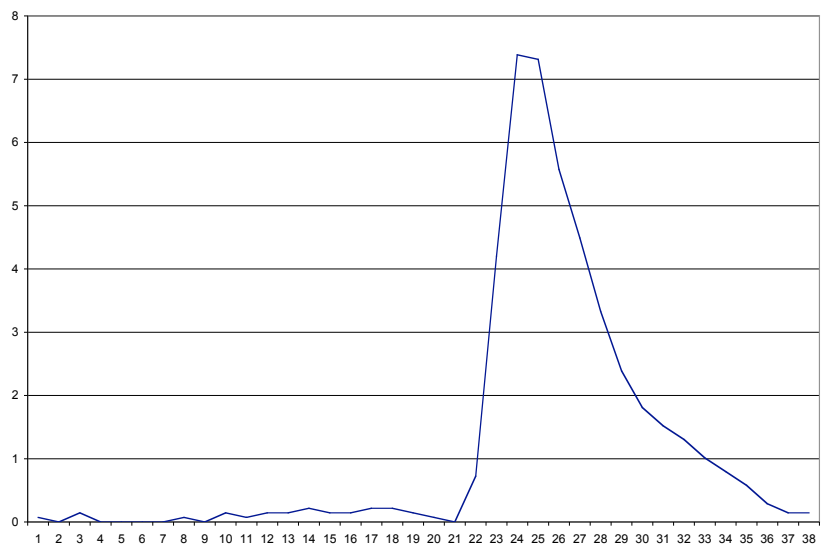

could be increased by as much as 3.5, reducing the calculated density by the same factor.)

Fig. 5 In the same shot, Shot 81602, near the peak of neutron power, 14 MeV neutrons are emitted from collisions of a beam of trapped tritium with deuterium (kW of DT neutron power). This occurs only 4 ns after the X-ray peak.

Given the above assumption, $n_i = 3.3 \pm 0.5 \times 10^{21}/cm^3$ for shot 81602 and $2.8 \pm 0.9 \times 10^{20}/cm^3$ for shot 80910. This is a **lower** limit, since it is based on the assumption that all the T has been produced in the plasmoid prior to the DT neutron pulse. If some of the T has been produced after the DT pulse, or outside the plasmoid in the plasma column, then F would be **larger** than the value cited here and n would be **larger**. This result represents a peak density. We take the average density to be half these values, $1.6 \pm 0.2 \times 10^{21}/cm^3$ for shot 81602 and $1.4 \pm 0.5 \times 10^{20}/cm^3$ for shot 80910.

We have ruled out any source for the DT neutrons other than a dense plasmoid where the tritons are produced, along with the DD neutrons. Collisions of tritons with deuterated metal in either the chamber wall or the anode, even assuming (unrealistically) that there is one deuterium atom for every copper or iron atom, would produce a burn fraction F of $<2.7 \times 10^{-5}$. This is a factor of 10 less than that observed for shot 80910 and a factor of 120 less than that observed for shot 81602. The tritons ejected in the ion beam will hit a copper target which is not in the detector line-of-sight, further reducing any possible beam-target contribution to the DT neutrons.

### 3.3 PLASMOID VOLUME AND CONFIRMATION OF DENSITY

Eric J. Lerner

Measurements of peak X-ray emitted power and peak $T_e$ (215keV) for shot 81602 give us that

$$1.6\times10^{-32}n_i^2VT_e^{1/2}=120\text{kW}$$

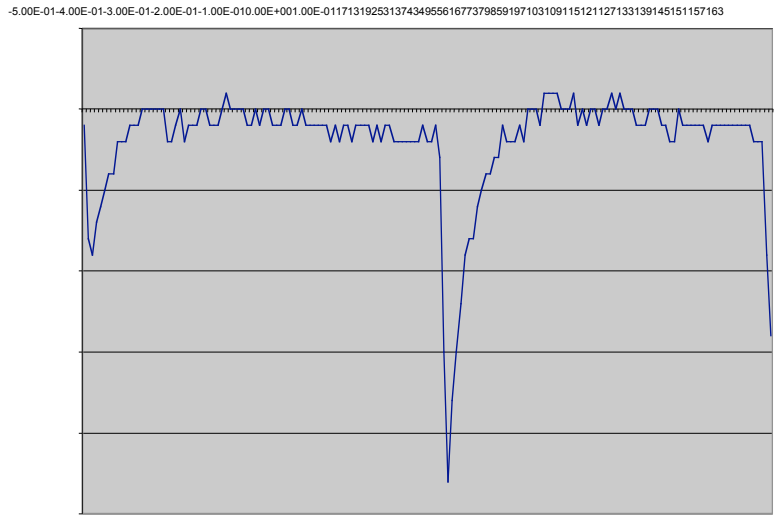

(a)

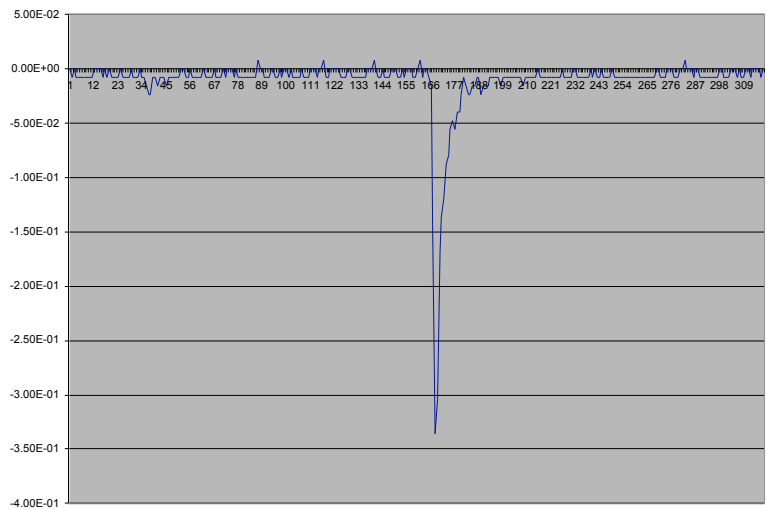

(b)

Fig. 6  Signal from 9 meter neutron detector(a) and 17.4 meter detector(b) for shot 81602, showing DT neutron pulse. Time is in 2 ns units and V is in volts. These unprocessed signal traces are scaled in time in proportion to the ratio of distance to the two detectors, so that particles traveling at a given velocity will appear in the same relative locations in the two traces. The central peak in (a) and the sole peak in (b) corresponds to 14.1 Mev DT neutrons. Note the high signal to noise ratios (25 to 1 in (a) and 40 to 1 in (b))and the lack of any confusing pulses. In (a) the earlier peak (to the left) is the gamma ray pulse from the e-beam and the rise to the right is the beginning of the much larger DD neutron peak, which begins in (b) a little to the right of the frame shown here.

$$n_i^2 V = 1.5 \times 10^{34}/cm^3$$

,where V is the volume of the reacting region.

Therefore, V is $5.5 \times 10^{-9}$ cm$^3$, indicating a core radius of 6 microns and an overall plasmoid radius of 24 microns. This is comparable to the dimensions of hotspots observed with pinhole X-ray cameras in many previous plasma focus experiments[6-8].

We did not attempt to directly measure the source size of the high energy X-rays in these experiments. There are difficulties with measuring the radius of the source of high-energy X-rays. Use of a pinhole camera is routine for low-energy X-rays. For a reasonable field of view, a pinhole must be cut in material that is thinner than the pinhole diameter. However, to have adequate contrast for X-ray energy above 100 keV, the lead thickness must be >200 microns, so resolution of tens of microns cannot be obtained in this manner. (After our experiment, Castillo-Mejia et al,[23] using an image intensifier, published results demonstrating that the source size of hard x-rays in their smaller PDF was of the order of 10 microns. In this experiment, however, they did not distinguish between X-rays from the plasmoids and gamma rays from the electron beam collision with the anode.)

For shot 81602, measurement of the emitted ion beam provides indirect confirmation of the density estimate. Measurements with a 10 cm diameter Rogowski coil located at 65 cm from the end of the anode showed that the ion beam had a peak current of 140 A and total ion number of $8.8 \times 10^{12}$. Since the number of ions in the plasmoids $n_i V$ must equal or exceed the number of ions in the beam, these measurements can combine with $n_i^2 V$ to set an upper limit on $n_i$ of $1.7 \times 10^{21}/cm^3$. This result is essentially identical to our estimate from DT neutrons, implying that essentially all the ions are evacuated into the beam.

This observed equality of the ion number in the plasmoid and in the ion beam allows an estimate of the density of other shots that did not yield DT neutrons. This is done by dividing the $n_i^2 V$ measured in each shot through X-ray or neutron observations by the beam ion number. This analysis indicates that $n_i$ ranges from $0.9$-$3.0 \times 10^{20}/cm^3$, with an average of $1.5 \times 10^{20}/cm^3$ for He shots and $2.5 \times 10^{20}/cm^3$ for D shots. These values are also consistent with the non-observation of DT neutrons in the other D shots. The factor of ten variations between best n and average n is typical of DPF functioning without extensive optimization.

The average electron density $n_e$ for the He and D shots was the same to within 20%. Since measurement of x-ray intensity allows the calculation for each shot of $n_e^2 V$, we find that V for the He and D shots are also closely similar. This in turn implies that the assumption that the efficiency of energy transfer from the electron beam to the plasmoid electrons is also similar for the two fill gases, made in Sec 3.1, is probably justified.

It may seem surprising or even incredible that 100keV deuterons can be contained for 50 ns in a volume only 12 microns across in the shortest dimension. The deuterons are thus trapped for about 13,000 crossing times. But the observation show clearly that 1.0 MeV tritons are trapped for at least 10 ns or about 6,700 crossing times, so the containment of the much less energetic deuterons does not thus appear so implausible.

It should be emphasized that the high densities measured in this experiment are quite comparable to those measured by a variety of techniques in other DPF research[11-13] and are not in themselves surprising. Many research teams have observed the same $10^{20}$-$10^{21}$ /cm$^3$ range with similar operating conditions. However, the present experiments are the first in which $T_e$ >100keV are associated with such high plasma densities.

At these high densities, electron-electron collisional heating is rapid, so there cannot be a cool background plasma with a high-energy component, meaning that essentially all the electrons must be energetic. For 200keV electrons with a density of $1.4 \times 10^{20}$ the electron heating time is only 6 ns, considerably shorter than the typical 20-50 ns x-ray pulse duration, and for $n = 1.6 \times 10^{21}$ (shot 81602) the heating time is only 0.5 ns. If we assume the hot electrons are kept hot by the beam, then enough energy would be transferred to the cold component to heat it up. To allow a hot and cold (20 keV) component to co-exist, the hot component must have an average electron energy > 1.3 MeV for $n = 1.4 \times 10^{20}$ and > 7 MeV for $n = 1.6 \times 10^{21}$. Such high energy for a hot component is completely ruled out by the x-ray absorption ratios observed.

Alternatively, can the cold electron component could be so much larger than the hot component that it heats up slowly? The electron population in the energy range 10-50keV, can be constrained directly because in some shots we removed the 300 micron copper filter from one detector. (X-rays shorter than 10 keV would be filtered out by the 170 cm of air in the line of sight to the DPF.) Comparing these observations with the 300 micron filtered ones, the excess x-rays at low energy are at most 50% of the hot (200kev) x-ray emission. This is consistent with an electron energy distribution somewhat broader than Maxwellian, but again rules out a large cold component. A 20 keV cold component must have a density $n_c < 3\, n_h$ to avoid violating these observational constraints.

But a cold component would have to be far more massive to avoid being heated up by the hot comment. In the case of shot 81602, where the cooling time of the 145Kev hot electrons in the presence of a cold compete is 80 times shorter than the x-ray pulse time, a cold component would have to have $n_c > 580\, n_h$ to avoid being heated beyond 20keV. Such a massive cool component can be rule d out if it has an average energy of 20keV or more. These x-ray observation cannot rule out an even colder component of 10keV or less, but such a colder component will be ruled out by arguments in sec. 3.4 .

## 3.4 ION ENERGY AND CONFINEMENT TIME

Not only is $T_e$ high in these plasmoids, but so is $T_i$, ranging from 45 to 210 keV. (These $T_i$ assume a Maxwellian distribution of ion energies. For monoenergetic ions, the range would be from 90 to 350 keV)[24]. For comparison, the ideal $T_i$ for hydrogen-boron fusion is 600 keV. For $T_i = 145$keV, the reaction rate is one third of its maximum. The highest $T_i$ achieved by TFTR was 44 keV [4].

$T_i$ and $\tau$ for the DD neutron production pulses were determined as follows. The FWHM of the neutron pulse was measured at 9 meters and at 17.4 meters. We made the simplifying assumption that the velocity distribution was Maxwellian. On that basis

$$1) \quad FWHM^2 = \tau^2 + (t^2 T_i / 2E),$$

where t is the transit time for a neutron of energy E to reach the detector and $T_i$ is the ion temperature, averaged over the duration of the pulse. (While it is unlikely the velocity distribution is in fact Maxwellian, monoenergetic ions would produce a different shaped neutron pulse than is observed, so a Maxwellian approximation appears roughly accurate.)

It is important to emphasize that the path between the plasmoids and the neutrons detectors was entirely clear of scattering walls, except for the 0.5 cm thick vacuum chamber wall. So there is no interference from the scattered tail of the neutrons, which is clearly distinguishable from the main peak.

For shot 81602, $T_i$ is 55 keV and ion confinement time $\tau$ is 54 ns. On this basis, $n\tau$ for this shot is $9.2 \times 10^{13}$ sec/cm$^3$ and $n\tau T_i$ is $5 \times 10^{15}$ keVsec/cm$^3$, a record value for any fusion experiment. Simultaneous X-ray observations show an average $T_e$ of 145 keV. For other shots, which have smaller n but larger $\tau$ and $T_i$, $n\tau T_i$ is typically $1 \times 10^{15}$ keVsec/cm$^3$.

There is abundant evidence that the high-energy electrons that produce the hard X-rays do in fact originate in the same dense hot spots that produce the neutrons. Therefore $T_i$, $T_e$ and n are all high in same time and space. The X-ray pulses are closely correlated in time and duration with the neutron pulses, overlapping with them in all cases, and with the X-ray peak preceding the center of the neutron pulse by only 5-18 ns (average 10 ns). The DT pulse peak for shot 81602 is within 4 ns of the X-ray pulse peak. The duration of the x-ray pulses and neutron pulses are closely correlated, with X-ray pulse duration, always 0.4-0.55 that of the neutron pulse. In addition, from the neutron production, pulse duration, and $T_i$, the quantity $n^2V$ was calculated for each deuterium shot using the standard formula

$$Y_n = n_i^2 V\tau \langle\sigma v\rangle.$$

The same quantity can be calculated from $T_e$ and X-ray power for the D shots with high X-ray signal to noise using

$$P_x = 1.6\times 10^{-32} n_i^2 V T_e^{1/2}$$

Where $P_x$ is average x-ray power. It is important to note that the quantity $n^2V$ calculated in this manner does not depended on knowledge of n or V separately. In the case of the x-ray data, both $P_x$ and $T_e$ come directly from observations, while for the neutron data $n_i^2V$ depends on the calculated values of t and $T_i$.

The average $n^2V$ derived from the neutron data is $1.7\times 10^{34}/\text{cm}^3$, while that derived from the X-ray data is $2.2\times 10^{34}/\text{cm}^3$, essentially identical. For shot 81602 the $n^2V$ of $1.5\times 10^{34}/\text{cm}^3$ derived from X-ray data is essentially identical to the $1.6\times 10^{34}/\text{cm}^3$ that can be derived from DD neutron emission, neutron pulse duration and ion energy $T_i$. These would be extraordinary coincidences if the neutrons and X-rays did not in fact originate from the same volume of plasma.

The high density-confinement-time product is important for fusion performance since it is directly proportional to the fraction of fusion fuel burned. For comparison the best $n\tau T_i$ achieved in the Tokamak Fusion Test Reactor was $5.5\times 10^{14}$ keVsec/cm$^3$ with $T_i$ of 44 keV and $T_e$ of only 11.5 keV[4].

Since this claim may seem sensational, given the far larger effort expended on the TFTR, it deserves some elaboration. First of all, the $\tau$ of 54 ns derived from time of flight neutron data can be independently confirmed from a different set of data for the same shot. As described in Sec.3.3, x-ray intensity measurements, together with measured $T_e$ yield $n_i^2V = 1.5\times 10^{34}/\text{cm}^3$. Since the number of DD neutrons $Y_n$ is $1.1\times 10^{10} = n_i^2 V\tau\langle\sigma v\rangle$, we have

$$2)\; \tau = Y_n/n_i^2 V\langle\sigma v\rangle = 58 \text{ ns},$$

essentially the identical result as the 54 ns derived from the time of flight data alone.

This good agreement is also a confirmation of the value of $T_i$, since $T_i$ enters into to the calculation of $\tau$ in different ways in eq.(1) and (2), where $\langle\sigma v\rangle$.depends on $T_i$ For example, if $T_i$ is arbitrarily assumed to be 27keV (half its calculated value of 55keV), then $\tau$ from TOF at via eq(1) become 83 ns and from x-ray measurements and $Y_n$ via eq (2) $\tau$ becomes 162 nsec, a very poor agreement. Indeed only for the range of $T_i$ from 55keV to 63 keV would the agreement be as good as the 7% difference found for the calculated value of 55 keV.

It could be argued that the time of flight data reflects bulk motion of the plasma, and the true ion energy available for collisions is thus far less than $T_i$. But in that case the above agreement would have to be viewed as an extraordinary coincidence.

It could be objected that $\tau$ is not a true ion confinement time, since the same observation would be obtained if a much large flow of ions through the plasmoid occurred, so that the confinement time of each individual ion was much shorter than $\tau$ and therefore the burn fraction smaller than that implied by the $n\tau$ quoted here. If a quantity of ions N>

$n_i$ V flowed through the plasmoid during the time $\tau$ ion confinement time would be reduced to $n_i V\tau/N$.

However, this would not affect the value of $n\tau$. The tritons generated by the DD reactions would also not be fully confined. At any given moment we would expect at most a fraction $n_iV/N$ of the tritons generated would be within the plasmoid. This assumption would in turn increase the burn fraction F, which was calculated on the assumption that all the tritium produced by the DD reactions was in the plasmoid at the same moment. If only a fraction $n_i V/N$ was present, the calculated n would be increased by a factor $N/n_i V$, and thus $n\tau$ would be unchanged. Our calculation of $n\tau$ and thus $n\tau T_i$ is a robust one.

Similar arguments apply if it is assumed that more than one plasmoid are contributing to neutron and x-ray production. If, say N plasmoids all contribute to the DT neutron production shown in Fig. 5 then they are simultaneous to within a few ns and could not masquerade as a much longer pulse, unless their lifetimes were also in the 50 ns range. If only one plasmoid produces the entire DT pulse, with other plasmoids contributing to the DD neutron production, then for the DT-producing plasmoid, n must be increased by a facet of N since only 1/N of the tritium is involved. Thus for five 10-ns plasmoids imitating a single 50 ns pulse, n is increased by five times and $\tau$ is decreased by the same factor gain leaving $n\tau$ unchanged.

If the $T_i$ derived from TOF data are accepted as the valid values, we can put firm limits on any cold component in the plasmoid, even one colder than 10keV. For shot 81602, $n_i^2 V$ =$1.5\times10^{34}$/cm$^3$, which means that $(n_c+n_h)^2 V=1.5\times10^{34}$/cm$^3$. But since from x-ray data, $n_h$ $(n_c+n_h)= 1.6\times10^{34}$/cm$^3$, essentially the identical figure, $n_c$ must be $<< n_h$. An $n_c>580\ n_h$, as required to avoid rapid heating, is clearly ruled out by this argument.

## 3.5 ENERGY CONSIDERATIONS

Since the ultimate aim of this research is to arrive at net energy production, it is important to consider the efficiency of energy transfer at various stages of the plasma focus process. To begin with the plasmoid itself, the minimum magnetic energy in the plasmoid can be determined from the observation that the 1.0 MeV tritons are trapped within the magnetic field for a period of 10 ns. Since the crossing time of the plasmoid for the tritons with $v=7.9\times10^8$ cm/s, was only 1.5 ps, magnetic confinement clearly occurred, which meant that the gryoradius of the tritons must have been less than the radius of the core of the plasmoid. Thus,

$$vMc/eB< r,$$

where M is the mass of the triton and v its velocity. This condition can only be met if the current in the plasmoid core $I_c> 1.2$ MA. For shot 81602, with a core radius of 6 microns and a core length of 48 microns (with the aspect ratio being set equal to that observed in many other experiments) minimum B is $4\times10^8$ G and minimum magnetic field energy is 3.5 J. The plasmoid current would then be essentially the same as the peak input current and the B field, while large, is within the range observed in several other DPF experiments[5].

The thermal energy of the ions and electrons combined can be calculated from $nVT_e$ and $T_i$ and is 0.3J assuming a Maxwellian distribution or 0.5J assuming monenergetic ions and electrons. Total plasmoid energy must therefore be >3.8J.

We can compare this energy with that emitted by the ion beam. Time of flight measurement using the two Rogowski coils show that ions in the beams have energies that are peaked consistently in the range of 1.1-1.5 MeV. In shot 81602 the beam had a peak current of 140 A, total power of 180 MW and total energy of 1.8 J. Since the electron beam, accelerated by the same field within the plasmoid, must have had the same energy initially (before some of its energy was transferred to the plasmoid electrons) total beam energy was 3.6 J.

Since this value is essentially identical with the 3.8 J minimum energy for the plasmoid the plasmoid energy appears to be transferred at high efficiency to the beams as predicted by the theoretical model used here[16,17]. This has to be a qualified conclusion since the plasmoid energy is a lower limit.

The energy of the electron beam can also be measured, although less directly, as a check on the ion beam measurement. The ~1.0 MeV gamma rays that form the initial pulse observed through the lead shields of the remote detectors clearly are generated by the electron beam hitting the copper anode. Taking an electron energy of 1.3MeV, equal to that measured in the ion beam, it takes a beam energy of 0.72 J to produce the observed 5.3 microjoule gamma ray pulse, observed through 5 cm lead, 6 cm copper (from the anode and the cover of the detector) and the 9mm steel vacuum chamber wall. Since 0.3-0.5 J of the initial electron beam energy is absorbed in heating the plasmoid particles, the initial e-beam energy is 1.0-1.2 J, in adequate agreement with the 1.7 J measured for the ion beam. At least 18% of the electron beam energy is transferred to the plasmoid electrons.

Total fusion energy released in shot 81602 is 11.7 mJ or 0.3% of total plasmoid energy and 4% of particle thermal energy. Total x-ray energy released is 2.2mJ or 20% of fusion yield.

Part of this same analysis can be done for shot 80910, which, however lacks x-ray data. Here minimum magnetic energy is 19J, minimum particle energy is 3.6 J, ion beam energy is 4.4 J, electron beam energy is 3.0J. The agreement between total e-beam energy, taken as 6.6J and ion beam energy is again adequate. But the total plasmoid energy is in only marginal agreement with total beam energy, being about a factor of 2 larger. At least 38% of electron beam energy is converted to particle energy. Fusion energy of 58mJ is 0.3% of plasmoid energy and 1.6% of particle energy.

It is notable that in both shots, the beam energy is closely comparable to the total plasmoid energy and that in both cases beam energy exceeds plasmoid particle energy, in accord with the model's assumption that the beams heat the particles. This would not be the case, however, if $T_i$ were in fact considerably less than the $T_i$ calculated from TOF data, or if a large cool electron component were present. In either of these cases, particle energy would considerably exceed beam energy. So the agreement shown here is another indication that the TOF calculation is reliable and no cold component is present.

Clearly the efficiency of energy transfer into the plasmoid is not high, since by the time of the pinch the capacitor bank has transferred 41kJ into the device. (This is measured by the voltage probe and primary Rogowski coil). So only 0.01-0.05% of total energy available is fed into the plasmoid in this shot. Ways to increase this efficiency are discussed in Sec. 6.

Overall energy efficiency should be viewed in the context of other fusion devices. Total bank energy is 120kJ, so "wall plug" efficiency is $4.8 \times 10^{-7}$ for shot 80910, but for the highest neutron yield ($3 \times 10^{11}$) with any electrodes and the same input energy, efficiency is $2.4 \times 10^{-6}$. For comparison, the best D-D yield with TFTR is 11.7kJ and the total power supply energy for one pulse is 3GJ, producing a wall plug efficiency of $3.9 \times 10^{-6}$[4].

## 4. DISCUSSION AND COMPARISONS WITH EARLIER RESULTS

Given the significance of the results, it is important to examine carefully whether there is an alternative explanation. In particular, is it at all possible that the high energy electrons and ions are in a more diffuse plasma than the dense plasmoid which clearly produces the DT neutrons?

This alternative hypothesis necessitates that some fraction of the DD neutrons are produced by the diffuse plasma, the rest by the dense plasmoid that also produces DT neutrons. If, say, 0.5 of the neutrons are produced by the plasmoid and 0.5 by the diffuse plasma, then the plasmoid can also contain only half the tritium, and n must then be twice the value calculated above, or $3.2\times10^{21}/$ cm$^3$. Assuming that the plasmoid exists only long enough to produce the DT neutron pulse, or 10 ns, n$\tau$ would then be $3.2\times10^{13}$, a factor of 3 less than that calculated above.

If, in this alternative we assume that the other half of the neutrons are produced in a more extended plasma that also produces the hard x-rays, we can extend the cooling time of the hot electrons to 25 ns if $n_i= 2.2\times10^{19}/$ cm$^3$. We can avoid heating the cool component above 10keV by limiting the hot electrons to $n_h= 1.5\times10^{18}/$ cm$^3$. To produce the observed quantity of hard x-rays, we then have an extended plasma volume of $6.6\times10^{-4}$ cm$^3$, $10^{15}$ hot electrons and total hot electron energy of 23 J. In the same volume, if we were to assume $T_i$ ==55keV, we would produce 38 times more neutrons than observed. So we have to also assume that the 55keV TOF energy is a bulk motion of trapped plasma and thermal energy per ion is only 8keV. Total particle energy will then be at least 150 J and total magnetic field plus particle energy at least 300 J, since the magnet energy of the containing field must at least equal total particle energy.

Two things are obvious about this alternative model. First, it is highly artificial with ad-hoc assumptions about density and particle energy required to conform to observation, generally a sign of an incorrect model. Second, the energy involved is 100 times more than the energy in the beams. So in this model, the beams could not possible be the source of energy for the ions and electrons despite the fact that the beams are closely synchronous with the x-ray and neutron emissions. In contrast the model used in this paper explains all the observed phenomena with no ad-hoc assumptions. It is thus a much better explanation of the data.

The plasmoid-based model used here makes quantitative predictions that are confirmed by the data, giving another reason for confidence in this interpretation. As described in previous papers[5,16,17]) the plasmoid begins to generate the beams when the synchrotron radiation from the highest-field region at the core of the plasmoid can escape the plasma. This sudden loss of energy from a localized spot at the center of the plasmoid causes a drop in the current and in the magnetic field, generating a strong electric field that accelerates the ion and electron beam. The synchrotron radiation can begin to escape when the gyrofrequency just exceeds $2\omega_p$, where $\omega_p$ is the plasma frequency. For shot 81602 the ratio of peak gyrofrequency to peak plasma frequency is 2.2, an excellent agreement with prediction.

It is important to note that the results presented here are consistent in many ways with earlier work. In particular the density of hot spots in the $10^{21}/$ cm$^3$ range has been reported by many authors[11,12,13,25,26] measuring density by diverse means, including the ones used here--the ratio of DT to DD neutrons. Brzosko and Nardi reported[25] n$\tau$ of $1.6\times10^{13}$ s/cm$^3$, very close to our average results of $1.6\times10^{13}$ s/cm$^3$. The same team measured densities as high as $5\times10^{21}$ /cm$^3$, somewhat above the best levels reached here. Magnetic fields of the order of the 400 MG reported here, up to 200 MG, have also been reported and measured previously[5].

What is new in these results is the high reported Te$_i$ and T$_i$. Others have reported T$_e$ of 7keV[27] and T$_i$ of 18keV[28], much lower than the values reported here. There are two main reasons for this difference. The first concerns the conditions in this experiment. The fill pressure was deliberately increased by decreasing the electrode length, which I anticipated on theoretical grounds [16,17] would increase n, thereby increase the coupling between the electron beam and the plasma electrons. This in turn would increase T$_e$ and T$_i$. The validity of this correlation of fill pressure with T$_e$ is demonstrated in a set of shots run with a longer (30 cm) electrode at 9 torr. Using the same x-ray technique, average T$_e$ of 34 keV was found, a factor of 5 less than the average T$_e$ of 175keV found at 13 torr using the shorter electrode. (In these shots, the radiation from the anode was not blocked, but there was a clear distinction in time between the first, gamma-ray, pulse and the second, hard x-ray, pulse. If there was a gamma-ray contribution to the second pulse, it would increase the apparent temperature. In this case, the drop in $T_e$ at the lower fill pressure would be even more pronounced than described here.)

By comparison, none of the experiments found in the literature that measured $T_e$ or $T_i$ ran with fill pressures as high as 13 torr. The measurement of $T_e$ of 7kev at 1 torr fill pressure, and $T_i$ of 18keV measurement at 8 torr is comparable with the present $T_e$ measurement of 34keV at 9 torr.

In addition, almost all experiments failed to distinguish physically between the anode gamma ray emission and the hard x-ray emission from the plasmoids. This in some case led to erroneous conclusions that the MeV electrons were trapped in the plasmoids and thus, at these high energies could co-exist with a cooler background. As well, the approach of many groups to ignore the importance of the plasmoid and assume a relatively homogenous plasma within the pinch led to problems in interpreting data and the rejection of high $T_e$ and $T_i$, since these could not be representative of the entire pinch column.

## 5. MAGNETIC FIELD EFFECT

The high particle energies achieved in these shots already match those needed for hydrogen- boron fusion. For the He shots, the atomic mass is close to the average atomic mass of p$^{11}$B, so it is reasonable to expect that these conditions can be matched with p$^{11}$B. However, while the experiments described above show clearly that the high particle energies required of proton boron fusion are attainable, there is another well-known challenge in achieving net energy from this fuel. Because of the $z^2$ dependence and boron's z of 5, bremsstrahlung x-ray radiation is enhanced and many analyses have indicated that fusion poorer can barely if at all exceed plasma cooling by bremsstrahlung[29]. If unavoidable, this situation would eliminate the heating of the plasma by the fusion produced alpha particles and would require that all the energy be recovered from the x-ray radiation.

However published analyses have overlooked an important physical effect that is especially relevant for the DPF. This effect, first pointed out by McNally[30], involves the reduction of energy transfer from the ions to the electrons in the presence of a strong magnetic field. This in turn reduce the electron temperature and thus the bremsstrahlung emission.

For ions colliding with electrons with gyrofrequency $\omega_g$, energy transfer drops rapidly for impact parameters $b > v_i/\omega_g$, where $v_i$ is ion velocity, since in that case the electron is accelerated at some times during the collision and decelerated at others, so average energy transfer is small. This means that the $b_{max}$ is less than the Debye length, $\lambda_D$ by a factor of $v_i \omega_p / v_{et} \omega_g$, where $\omega_p$ is the plasma frequency and $v_{et}$ is the electron thermal velocity. So the Coulomb logarithm in the standard energy-loss formula is reduced to $\text{Ln}(mv_i^2/\_\omega_g)$.

This formula is approximately valid for collisions in which ions collide with slower moving electrons, which are the only collisions in which the ions lose energy. But for collisions of faster moving electrons with ions, where the electrons lose energy to the ions, the Coulombs logarithm, by the same logic, is $\text{Ln}(mv_{et}^2/\_\omega_g)$. If $v_{et} >> v_i$ then $\text{Ln}(mv_e^2/\_\omega_g)$ can be much larger than $\text{Ln}(mv_i^2/\_\omega_g)$ for sufficiently large values of $\_\omega_g$, in other words for sufficiently large B. Ignoring momentum transfer parallel to field, steady state occurs when $T_i/T_e = \text{Ln}(mv_e^2/\_\omega_g)/\text{Ln}(mv_i^2/\_\omega_g)$[30].

This effect has been studied in a few cases for fusion plasmas with relatively weak fields, where is shown to be a relatively small effect[31]. It has been studied much more extensively in the case of neutron starts[32],[33]. However, it has never been applied to the DPF plasmoids, whose force-free configuration and very strong magnetic fields make the effect far more important. First, small-angle momentum transfer parallel to the field can be neglected in these plasmoids, since the ion velocity lies very close to the local magnetic field direction, and $\Delta p_{par}/\Delta p_{perp} \sim \sin^2\theta$, where $\theta$ is the angle between the ion velocity and

Eric J. Lerner

the B field direction[32].

In a force-free configuration, such as the toroidal vortices that make up the plasmoids, ions disturbed by collisions return to the local field lines in times of order $1/\omega_{gi}$, so

$$\theta \sim \omega_{ci}/\omega_{gi},$$

Where $\omega_{ci}$ is the ion collision frequency. For a decaborane plasma, $\theta \sim 2 \times 10^{-8} n/T_i^{3/2} B$. For the example of the plasmoid conditions calculated above, $n=3 \times 10^{21}$, $B=400$MG, $\theta=0.01$ for $T_i=60$keV. In all plasmoids of interest in the DPF, $\theta$ is very small and small-angle parallel momentum transfer is thus also very small.

Even more significantly, the high B in plasmoid generates a regime where $mv_i^2/\_\omega_g<1$. In this case the magnetic effect is very large, the above formulae break down and quantum effects have to be considered. Such a situation has never been studied before for fusion applications, but has been analyzed extensively in the case of protons falling onto neutron stars[32].

In a strong magnetic field, since angular momentum is quantized in units of $\_$, electrons can have only discrete energy levels, termed Landau levels (ignoring motion parallel to the magnetic field):

$$E_b = (n+1/2) e\_B/mc = (n+1/2)\, 11.6 eV B(GG).$$

Viewed another way, electrons cannot have gyroradii smaller than their DeBroglie wavelength. Since maximum momentum transfer is $mv$, where $v$ is relative velocity, for $mv^2/2 < E_b$ almost no excitation of electrons to the next Landau level can occur, so very little energy can be transferred to the electrons in such collisions. Again ignoring the electron's own motion along the field lines, thus condition occurs when

$$E_i < (M/m)\, E_b$$

For $E_i = 300$keV, this implies $B>14$GG for p, $B>3.5$GG for $\alpha$, and $B>1.3$GG for $^{11}$B. As will be shown below, such field strengths should be attainable with the DPF.

If we assume that $T_{eth} >> E_b$, then we have to consider the motion of the electrons along the field lines, which can increase the relative velocity of collision, v. In the classical case, the ions will lose energy only from electrons for which $v_{epar} < v_i$. Since for these collisions $v < 2v_i$, energy loss will still be very small if $E_i < 1/2(M/m)\, E_b$, which can occur for boron nuclei.

However, there is a phenomena which prevents energy loss to the electrons from falling to negligble levels. (This were not considered in the presentation actually given at the symposium so the conclusions here differ somewhat from those presented then.)

In the classical case, considering only motion along the line of force, an ion colliding with a faster moving electron will lose energy if the electrons' velocity is opposite to the of the ion, but will gain energy if they are in the same direction--the electron overtaking the ion. In the latter case the relative velocity is less than in the former case, and since the energy transfer increases with decreasing relative velocity, there is a net gain of energy to the ion. For an ion moving faster than the electron, the ion overtakes the electrons, and thus loses energy independently of the direction that the electron is moving in. Thus ions only lose energy to electrons moving more slowly than they are.

However in the situation considered here, ions in some cases can lose energy to electrons that are moving faster than the ions. Consider the case of ions moving along the filed lines colliding with electrons in the ground Landau level. If $v_{epar}$ is such that $m(v_i+v_{epar})^2 > 2 E_b$, while $m(v_i-v_{epar})^2 < 2E_b$, the energy lost by the ion in collision with opposite-directed electrons will much exceed that gained in same-directed collisions, since in the first case the electron can be excited to a higher Landau level, but not in the second

Prospects for P$^{11}$B Fusion with the Dense Plasma Focus: New Resultscase. In neither case can the electron give up to the ion energy from perpendicular motion, as it is in the ground sate. (So, this consideration does not apply to above-ground-state electrons, which will lose energy to slower-moving ions.)

If $T_{eth} \gg E_b$, and we assume a Maxwellian distribution, the number of electrons in the ground state will be proportional to the volume in velocity-space. The number of such electrons in the ground state will be $\sim \pi v_i E_b$, while the number of electrons moving slower than $v_i$ in a non-magnetic Maxwellian distribution will be $\sim (8/3)\pi v_i E_i(m/M)$, so there will be a factor of $(3/8) E_i /E_i (m/M)$ increasing the effective collision rate, comparing the magnetic with non-magnetic case, for $E_i < E_b$.

At the same time, the rate of energy loss to each electron will be much less at relative velocities of order $v_b=(2E_b/m)^{1/2}$, as compared to relative velocities of order $v_i$ in the non-magnetic case. This *reduces* the energy loss for the magnetic case by a factor of $E_i(m/M)/E_b$. Combining these two factors, it is clear that the Coulomb logarithm term in the magnetic case tends to a constant value, independent of $E_i$ for $E_i < E_b$.

The calculation of the exact value of $\ln\Lambda$ for a given Maxwellian distribution of ions with dimensionless temperature $T=T_i/E_b(M/m)$, begins with a quantum mechanical calculation of $\ln\Lambda'$ for an ion moving along the field line colliding with an electron assumed to be at rest. This result was calculated by Nelson, Saltpeter and Wasserman [32]. From their work,

$$\ln\Lambda'(v) = (1/2)v(1/(2v^2) - e^{2v^2}\Gamma(0,2v^2))$$

Where $v$ is the dimensionless ion velocity $v_i/v_b$, and $v<1$. For $v>1$, the same source gives as a good approximation

$$\ln\Lambda'(v) = \ln(2v^2)$$

To determine the effective $\ln\Lambda''(v)$ for an ion of velocity $v$ colliding with ground state electrons with temperature $T_{eth} \gg E_b$, we can use these formulae, substituting in $v'$ as the relative velocity. First consider the case where $v<1$ and $v'<1$. There are equal number of electrons moving with positive or negative velocity relative to $v$, so we integrate assuming both $v'=v+v_e$ and $v'=v-v_e$, where $v_e$ is $v_e/v_b$. For $v>v_e$, both cases contribute to ion energy loss, while for $v<v_e$, the $v+v_e$ case adds to ion energy loss, while the $v-v_e$ cases subtracts. Taking into account the $1/v^2$ dependence of energy loss rates, we have the contribution to $\ln\Lambda''(v)$ from $v'<1$ collisions is

$$3/8\, v^{-1}\left(\int_0^v (v^{-2})\ln\Lambda'(v-v_e)\,dv_e + \int_0^{1-v}(v^{-2})\ln\Lambda'(v+v_e)\,dv_e - \int_v^{1+v}(v^{-2})\ln\Lambda'(v_e-v)\,dv_e\right)$$

Numerical integration shows that this integral is 0 for all $v'<1$. While individual collisions can add or subtract energy, collectively there is no net energy transfer when the relative velocity is too low to excite the electron out of the ground state, as would be naively expected.

Eric J. Lerner

This leaves the cases where $v'>1$. If $v+v_e >1+2v$, then $v_e-v >1$ as well, and in these cases since $v_e>v$ (still considering $v<1$) the ions gain energy on net. So additional energy loss can only come when $1< v+v_e <1+2v$. So the additional term (which contributes the whole of $\ln\Lambda''(v)$) is

$$\ln\Lambda''(v) = 3/8 v^{-1} \int_{1}^{1+2v} \ln(2v'^2) v'^2 dv' = 3/8 v^{-1} ((2+\ln2/2)(1-1/1+2v) - \ln(1+2v)/1+2v)$$

This expression is close to 3/8 for $0.2<v<0.5$ and decreases to $3/8\ln2$ as $v$ approaches 0 and to $\sim(3/8) \times 0.83$ for $v=1$.

For $v>1$, another term is needed to account for the case where $v>v_e$ and $v_e-v >1$, where energy is again lost to the electrons. In addition, for $v>(3/2)^{1/2}$, energy can be lost to non-ground-state electrons as well, and $\ln\Lambda''(v)$ rapidly converges on $((v-1)/v)\ln(2v^2)$. We can use this value for $v>1.35$.

Finally, integration over a Maxwellian distribution of $v$ yields the effective Coulomb logarithm as a function of T, the dimensionless ion temperature. (As a first approximation, the use of Maxwellian distributions is justified by the fact that the ion-ion and electron-electron energy transfer rates considerably exceed the ion-electron rates. However, the ion velocity distribution will be distorted by the alpha-particle heating.)

$$\ln\Lambda(T) = 3/8 \int_{0}^{1.35} (e^{-(v^2/T)}/v^2)((2+\ln2/2)(1-1/1+2v)-\ln(1+2v)/1+2v) dv$$

$$+3/8 \int_{1}^{1.35} (e^{-(v^2/T)}/v^3)((2+\ln2/2)(v-1)-2\ln v) dv$$

$$+ \int_{1.35}^{\infty} (e^{-(v^2/T)}/v)((v-1)/v) \ln(2v^2) dv$$

This result is presented in Table 2.

Table 2

| T | $\ln\Lambda(T)$ |
|---|---|
| 0.05 | .346 |
| 0.1 | .353 |
| 0.2 | .354 |
| 0.3 | .350 |
| 0.4 | .349 |
| 0.5 | .350 |
| 0.6 | .353 |
| 0.8 | .368 |
| 1.0 | .392 |
| 2.0 | .567 |
| 3.0 | .755 |
| 4.0 | .926 |
| 6.0 | 1.218 |

For the heating of the ions by the much faster thermal electrons, with $T_e>>1$, quantum effects can be ignored and the coulomb logarithm is simply $Ln(2T_e)$.

## 5.1 CALCULATION OF ELECTRON TEMPERATURE, BREMSSTRAHLUNG

Obtaining the Coulomb logarithm allows the calculation of energy balances for any combination of T, $n\tau$, and B. To illustrate the magnitude of the magnetic effect and to demonstrate that fusion power can well exceed bremsstrahlung loses, I calculated a number of examples, using the theoretical model of the plasma focus described in section 6. In these examples $T_i$, the ion temperature is assumed to be 300keV, the B and $n\tau$, are calculated based on the formulae in section 6. In these calculations $T_e$ is determined by finding the value at which no net energy is flowing to the electrons. In this case the net energy loss through the heating of the protons and $^{11}$B and through bremsstrahlung radiation is balanced by the heating of the electrons by fusion-generated alpha particles and the heating by the electron beam, whose energy is assumed to be wholly absorbed. The fuel is assumed to be pure decaborane($B_{10}H_{14}$).

The examples chosen are for very small electrodes with a cathode radius of only 1.25 cm. However, very similar results are obtained with electrodes with a cathode radii of 5 cm and peak currents just double those given in table 3 (magnetic fields half as big but the same $n\tau$).

Table 3

| Peak Current (MA) | B (GG) | $n\tau$ ($10^{15}$sec/cm$^3$) | $T_e$ (keV) | Fusion/ brem |
|---|---|---|---|---|
| 1.5 | 24 | 6.0 | 24 | 2.1 |
| 1.4 | 22.4 | 5.2 | 28 | 1.9 |
| 1.3 | 20.8 | 4.5 | 34 | 1.8 |
| 1.2 | 19.2 | 3.8 | 47 | 1.5 |
| 1.1 | 17.6 | 3.2 | 75 | 1.2 |

Note that the effect is sensitively depend on $n\tau$, since as n decreases collisional processes become less important relative to beam heating. For comparison, if magnetic effects are entirely neglected, at 1.5 MA $T_e$ is 113keV and Fusion/ bremsstrahlung power is .97. In contrast, with the correct accounting for magnetic field effects fusion energy release is well above bremsstrahlung, leading to rapid heating of the ions.

This calculation is of course only a preliminary one for a specific static case. A dynamic model that follows a plasmoid as it heats up and as helium ash is accumulated will be a logical next step in this analysis. But the analysis presented here clearly shows that net energy production is possible with decaborane at high magnetic fields.

## 6. CAN THESE CONDITIONS BE ACHIEVED WITH THE DPF?

Previously-developed theory can show how these conditions can be achieved with a

Eric J. Lerner

DPF. As described in [16,17] the DPF process can be described quantitatively using only a few basic assumptions. First, we assume that the magnetic energy of the field is conserved during the formation of the plasmoid, and that in a well-formed pinch, all the energy present in the field at the time of the pinch is trapped in the plasmoid. Given that experimentally, it has been determined that the length of the central channel in the plasmoid is close to 8 times its radius, we have:

$$I_c^2 r_c = I^2 r/8$$

where $I_c$ is current(A) in the plasmoid, $r_c$ is the radius(cm) of the central plasmoid channel, I is current at time of pinch and r is the cathode radius. (The cathode is the outer electrode.)

Second, following[16], plasma instability theory shows that for optimal filament formation, in the plasma chamber,

$$\omega_{ce} = \omega_{pi}$$

where $\omega_{ce}$ is electron gyrofrequency and $\omega_{pi}$ is ion plasma frequency. This immediately allows us to predict the optimal pressures given r and I, the plasma velocity, and thus the electrode length for a given pulse length.

$$n_i = (\mu M/m) I^2 / 100\pi mc^2 r^2$$
$$V = c(m/\mu M)(r/R),$$

where $n_i$ is initial ion density, $\mu$ is atomic mass, V is the peak sheath velocity at the anode, R is the anode radius, m is electron mass and M is proton mass.

Third, instability theory can also be used, as in [16]. to show that in the filaments

$$\omega_{cef} = \omega_{pep}$$

where $\omega_{cef}$ is the electron gyrofrequency in the filament and $\omega_{pep}$ is the electron plasma frequency in the background plasma.

We can then find that the incoming filament system, and thus the DPF as a whole has an effective resistance of

$$15/(\mu M/m)^{3/4} \text{ ohms}$$

so that the peak I for a given V can be determined. It should be noted that this is a maximum value, and that it can only be obtained if the inductance of the pulsed power supply plus DPF is sufficiently low.

Fourth, we know that at the time the plasmoid begins to decay,

$$\omega_{ce} = 2\omega_{pe}.$$

This is due to the condition that when the synchrotron frequency exceeds twice the plasma frequency, energy can be radiated. At this point, the current begins to drop, and the change in the magnetic field sets up large accelerating potentials to sustain the current. This in turn generates the ion and electron beams that release the energy trapped in the plasmoid and initiate its decay, as well as start nuclear reactions.

Finally, we assume that during compression the ratio B/n is a constant, as explained in

[16].

From these basic physical relations, it is simple algebra to derive the plasma parameters in the plasmoid, not only for hydrogen, as in [16], but for any gas or mixture of gases[17]. The results are summarized here:

$$3) \; r_c = 2^{-7/3}\mu^{-2/3}z^{-2/3}M/m^{-2/3}r = 1.32 \times 10^{-3}\mu^{-2/3}z^{-2/3}r$$

$$4) \; B_c = 4z(\mu M/m)B$$

$$5) \; n_c = 3.7 \times 10^{10}\mu^2 z I^2/r^2$$

where $B_c$ is peak field at cathode. The model in [16] also allows us to describe the production of the electron and ion beams and the duration of the plasmoid. This is possible simply by equating the energy lost though the beams with the decay of the plasmoid B field, allowing a calculation of the accelerating potential, beam current and decay time.

$$6) \; \tau = 6.2 \times 10^{-6} r_c/R_B = 8.2 \times 10^{-9} \mu^{-2/3} z^{-2/3} r/R_B$$

$$7) \; n\tau = 304 \mu^{4/3} z^{1/3} I^2/rR_B$$

$$8) \; E_b = I_c R_B e/4\pi^2 = .24 \mu^{1/3} z^{1/3} I R_B$$

$$9) \; I_b = .24 \mu^{1/3} z^{1/3} I$$

Here, $\tau$ is plasmoid decay time, $R_B$ is the effective resistance of the beam, $n_c$ is plasmoid density, $E_b$ is average beam energy per electronic charge and $I_b$ is beam current. However, a modification must be imposed here. For low I and thus low accelerating potentials, all the particles in the plasmoid are evacuated through the beam without carrying all the energy away. In this case the simple model will break down near the end of the plasmoid decay. However, for present purposes a suitable approximation simply reduces the plasma lifetime by the ratio of the accelerating potential to that needed to carry the entire plasmoid energy. To a good approximation this factor turns out to be I/1.4MA. For I>1.4MA, this factor is unity.

The particle density increases with $\mu$ and z as well as with I, and decreases with increasing r. Physically this is a direct result of the greater compression ratio that occurs with heavier gases, as is clear from the above relations. We thus see that the crucial plasma parameter $n\tau$ improves with heavier gases. Indeed, this is a faster improvement than appears at first, since it can be shown that $R_B$ also decreases as $\mu^{-3/4}$. Assuming z and $\mu$ to be proportional, $n\tau$ thus increases as $\mu^{2.4}$.

These theoretical precautions are in good agreement with the results discussed in sections 1-4. . If we use (4) to predict $B_c$ we obtain 0.43 GG, in excellent agreement with the observed value of 0.4 GG. Similarly, (7) yields $4.6 \times 10^{13}$ sec/cm$^3$ as compared with the best observed value of $9 \times 10^{13}$ and the average of $0.9 \times 10^{13}$.

For decaborane with z=2.66 and u =5.166, with r= 1.25cm and I= 1.5 MA, the formulae above yield $B_c$ =24GG and $n\tau = 6 \times 10^{15}$, while r= 5 cm, I =3MA yield B =12GG and nt =$6 \times 10^{15}$. This is of course a considerable extrapolation-- a factor of 60 above our observed value in both B and $n\tau$. However, these conditions can be reached with relatively small plasma focus devices.

These results show the potential of the DPF as a possible device for advanced fuel

Eric J. Lerner

fusion. Further work will be required to increase the efficiency of energy transfer into the plasmoid, which is low in these shots. Such high efficiency has been achieved with other DPF experiments[33,34,35,36], but not with the high density or particle energy achieved in these experiments. However, optimization of electrode design, which we were not able to accomplish in this experiment, may solve this problem. Our results show that neutron yield increases rapidly with increased sheath run-down velocity, in accord with other published results[37].This is confirmed by several other experiments[33,34,35,36]. The efficiency of energy transfer from the magnetic field into the plasmoid can be measured by comparing total beam energy with magnetic field energy at time of peak current. In a number of experiments where this efficiency is in the range of 10-20%, the Alfven velocity, $V_A$ at the anode at peak current is 18-23 cm/µs. In contrast, in our experiments, $V_A$ is 7.9cm/µs for D and 7.1 cm/µs for $^4$He and efficiency is only 0.01%. Therefore, a smaller anode radius, which will increase the B field and run-down velocity, could increase efficiency to much higher levels. A combination of high fill pressure, small anode radius, and high rundown velocity should be able to simultaneously achieve high efficiency of energy transfer as well as the high density and particle energy needed for advanced fuel fusion. $V_A$ in the region of 30 cm/µs for D is in the likely range of parameters needed for high efficiency and high $T_e$, $T_i$ and n. The theoretical model indicates that $V_A$ of about 15 cm/µs will be required for $p^{11}B$, but further experimental work will be needed to confirm this.

This research was funded in part by Jet Propulsion Laboratory and the Texas Engineering Experiment Station. The author thanks Bruce Freeman, Hank Oona, Alvin D. Luginbill, John C. Boydston, Jim M. Ferguson, and Brent A. Lindeburg for their participation in the experiments and Paul Straight for his technical assistance. The author also thanks Paul Finman for a request that led to the work in section 5.